\documentclass[aps,prb,floats,superscriptaddress,twocolumn,nobibnotes]{revtex4}
\usepackage[dvips]{graphicx}
\usepackage{amsmath}

\draft

\newcommand{\scrscr}{\scriptscriptstyle}

\newcommand{\la}{\langle}
\newcommand{\las}{\langle\!\langle\:}
\newcommand{\ra}{\rangle}
\newcommand{\ras}{\:\rangle\!\rangle}

\newcommand{\ce}[1]{\mbox{$c_{#1\sigma}^{}$}}
\newcommand{\cek}[1]{\mbox{$c_{#1\sigma}^{\dagger}$}}
\newcommand{\eSz}[1]{\mbox{$S_{#1}^z$}}
\newcommand{\eSS}[2]{\mbox{$S_{#1}^{#2\sigma}$}}
\newcommand{\en}[1]{\mbox{$n_{#1\sigma}^{}$}}
\newcommand{\Te}[1]{\mbox{$T_{#1}^{}$}}
\newcommand{\de}[1]{\mbox{$\delta_{#1}^{}$}}
\newcommand{\zet}[1]{\mbox{$z_{#1\sigma}^{}$}}

\newcommand{\Geup}[1]{\mbox{$G_{#1\uparrow}^{}$}}
\newcommand{\GAup}[1]{\mbox{$\Gamma_{#1\uparrow}^{}$}}
\newcommand{\GeupR}[1]{\mbox{$\stackrel{{}\, \scrscr R}{G}_{#1\uparrow}^{}$}}
\newcommand{\GOR}[1]{\mbox{$\stackrel{{}\hspace{.4em}\scrscr R}{\stackrel{}{\stackrel{}{}}} \hspace{-1em}{G}_{#1}^{(0)}$}}
\newcommand{\GAupR}[1]{\mbox{$\stackrel{{}\, \scrscr R}{\Gamma}_{#1\uparrow}^{}$}}

\newcommand{\ceup}[1]{\mbox{$c_{#1\uparrow}^{}$}}

\newcommand{\cekup}[1]{\mbox{$c_{#1\uparrow}^{\dagger}$}}

\newcommand{\ceupT}[1]{\mbox{$\tilde{c}_{#1\uparrow}^{}$}}
\newcommand{\ceupR}[1]{\mbox{$\stackrel{{}\, \scrscr R}{c}_{#1\uparrow}^{}$}}

\newcommand{\GFjup}[1]{\mbox{$\las #1 \,;\, \cekup{j} \ras$}}

\begin{document} 
 
\title{Local Density of States in the Antiferromagnetic and Ferromagnetic 
Kondo Models}

\date{\today}  
 
\author{P.~Sinjukow}
\email{peter.sinjukow@physik.hu-berlin.de}
\affiliation{Lehrstuhl Festk{\"o}rpertheorie, Institut f{\"u}r Physik,
  Humboldt-Universit{\"a}t zu Berlin, Invalidenstr.\ 110, 10115 Berlin,
  Germany}
\author{D.~Meyer} 
\affiliation{Lehrstuhl Festk{\"o}rpertheorie, Institut f{\"u}r Physik,
  Humboldt-Universit{\"a}t zu Berlin, Invalidenstr.\ 110, 10115 Berlin,
  Germany}
\affiliation{Department of Mathematics, Imperial College, 180 Queen's Gate, 
   London SW7 2BZ, UK}
\author{W.~Nolting}
\affiliation{Lehrstuhl Festk{\"o}rpertheorie, Institut f{\"u}r Physik,
  Humboldt-Universit{\"a}t zu Berlin, Invalidenstr.\ 110, 10115 Berlin,
  Germany}

\begin{abstract}
Based on a simple approximation scheme we have 
computed the local density of states (LDOS) of the antiferromagnetic and
ferromagnetic Kondo models
for the full range of
band occupations and coupling strengths. 
For both models 
the LDOS 
with its full energy dependence has not been
calculated before.
Arguments 
are 
given for the 
results 
to be 
qualitatively trustworthy
despite the simplicity of the approximation scheme.

\end{abstract}


\maketitle

\section{Introduction}

The term 
{\it antiferromagnetic Kondo model} 
stands for the well-known single-impurity {\it Kondo model},
whose origin is ascribed to Zener \cite{Zen51}. 
The attribute {\it antiferromagnetic} is used to 
stress 
the difference to 
the {\it ferromagnetic Kondo model}. 
In both models 
the spin of a magnetic
impurity is coupled to the spins of the conduction electrons
of a non-magnetic host lattice. 
In the (antiferromagnetic) Kondo model the sign of the coupling
constant is such that
antiparallel coupling between the
impurity and the conduction-electron spins is favoured (antiferromagnetic intraatomic
exchange). In the ferromagnetic 
Kondo model,
with the opposite sign of the coupling constant, parallel coupling is
favoured (ferromagnetic intraatomic exchange).  

The (antiferromagnetic) Kondo model has been one of the most intensively
discussed many-body 
mod{\nolinebreak}els in solid state physics since 
Kondo in 1964
succeeded in
explaining the resistance minimum of metals with small amounts of 
transition 
element impurities \cite{Kon64}.  
Thorough theoretical investigations 
later on led to the discovery of the cause of the 
minimum, 
the 
Kondo effect,
which is the collective (many-body) screening of the impurity spin by the spins
of the conduction electrons below the 
Kondo temperature $T_K$ \cite{TsW83,Hew}. 
By means of Wilson's numerical renormalization group 
theory (NRG) \cite{Wil75}  
and Bethe-ansatz methods \cite{Andr80,Wie80} exact 
solutions 
for 
thermodynamic properties such as the magnetic susceptibility
and the heat capacity were obtained.
Later the NRG was successfully extended to 
the computation of dynamic
(energy-dependent) quantities \cite{FrO86,SSK89,CHZ94}.

Although the antiferromagnetic Kondo model can 
thus 
be regarded as in principle solved, 
its local density of states (LDOS)
with the full energy dependence 
has not been calculated so far \cite{xxx}.
Nevertheless the LDOS is an interesting quantity, particularly since
scanning tunneling spectroscopy 
(STS) has made it experimentally accessible
\cite{CLE93,LSBD98,MCJCW98,MLE00}.

The LDOS of the ferromagnetic Kondo model, too,
has not been calculated yet. 
The ferromagnetic Kondo model 
has generally received less attention than 
its antiferromagnetic counterpart. This is probably due to 
the fact that 
a spectacular ``ferromagnetic Kondo effect''
does not exist 
because the effective coupling 
does 
not 
scale to infinity as in the antiferromagnetic Kondo model \cite{TsW83,Hew}. 
A second reason might be that there are fewer experimental realizations
than in the antiferromagnetic case.

In this 
paper 
we present 
results for the $T=0$-LDOS of both the antiferromagnetic and the
ferromagnetic Kondo models 
for the full 
range of band occupations and coupling strengths, gained 
by means of a simple 
Green's function approximation scheme based on
Nagaoka's decoupling procedure \cite{Nag65}.
The equations of Nagaoka's decoupling scheme were solved analytically by
Zittartz and M\"uller-Hartmann \cite{ZiM68}. However, their
equations still depend self-consistently on the scattering matrix.
Therefore, 
in order to find self-consistent numerical 
solutions,
it is equally well justified 
to use Nagaoka's original equations.

Qualitatively results 
prove trustworthy 
despite the crudness of the approximations.
For weak couplings the main correlation feature 
in the impurity-site LDOS of the antiferromagnetic Kondo model 
is an antiresonance, which is slightly shifted with
respect to the Fermi level.
There is an analogous antiresonance 
in the LDOS 
of the single-impurity Anderson model. A related
antiresonance structure was observed in recent STS measurements on magnetic
impurity adatoms 
\cite{LSBD98,MCJCW98,MLE00}.
For strong couplings the dominant characteristic
are quasiparticle $\delta$-peaks.
These can be traced back to 
the
limiting case of an empty conduction band, for which 
exact results and rigorous interpretation are given, 
and the atomic limit \cite{NoM84}. 

In the ferromagnetic Kondo model the LDOS at the Fermi level is practically
independent on the coupling strength, which is consistent with the fact
of a vanishing phase shift. 
However, the LDOS is diminished 
in the vicinity, 
giving rise to a peak-like structure.
If representing a true feature,
this peak structure should be measurable in STS experiments on magnetic
adatoms with \textit{ferromagnetic} intraatomic exchange.

It is clear that the approximation scheme is too simple to 
yield quantitatively reliable results.
Subtle features like the exponential Kondo scale in the
antiferromagnetic Kondo model cannot be reproduced.
Much better methods 
like numerical renormalization group theory (NRG) or Quantum Monte Carlo
(QMC) \cite{HiF86} 
would be 
required
for that.

This paper is structured as follows: 
First the exact solution for 
the limiting case of an empty conduction band
(Sec.~\ref{subsec:Theory_ngl0}) and the approximation scheme for finite
band occupation (\ref{sec:Appr_scheme}) are explained. 
Then the results for the LDOS are discussed in detail: first, for the
case of an empty conduction band (\ref{subsec:empty_band}), 
which is similar in both models, second, 
for the case of finite band occupation in the antiferromagnetic Kondo model 
(\ref{sec:afmKM}), and third, 
for finite band occupations in the ferromagnetic Kondo model
(\ref{sec:fmKM}). 
After that the influence of the magnetic impurity on
occupation numbers is 
discussed 
with emphasis on a seemingly contradictory result by other authors 
(\ref{sec:occup_numbers}). 
Finally we draw some conclusions (\ref{sec:conclusions}).

\section{Theory}
\label{sec:Theory}

The Hamiltonian of the antiferromagnetic and ferromagnetic Kondo models is given by 
\begin{eqnarray}
{\cal H}
=
\sum_{ij\sigma} \Te{ij} \cek{i} \ce{j} - J \,{\bf S}_0 \cdot 
{\mbox{\boldmath${\sigma}\!{}$\unboldmath}\,}_0
\label{eq:H-op} \;\;.
\end{eqnarray}
Replacing ${\mbox{\boldmath${\sigma}\!$\unboldmath}\,}_0$
by Fermi operators 
yields 
\begin{align}
{\cal H}=
\sum_{ij\sigma} \Te{ij} \cek{i} \ce{j}
- \frac{J}{2} \sum_{\sigma} (\zet{} \eSz{0} 
\en{0} + \eSS{0}{} \cek{0-} \ce{0}) 
\label{eq:Ham-op_ausf}
\end{align}
with $\zet{}=\de{\sigma\uparrow}-\de{\sigma\downarrow}$
and $\eSS{0}{} = \de{\sigma\uparrow} S_{0}^{+}
+\de{\sigma\downarrow} S_{0}^{-}$.
The first part of ${\cal H}$ with the hopping integrals 
$\Te{ij}$ describes the free movement of 
conduction electrons in a nondegenerate band.
$c_{i\sigma}^{(\dagger)}$ 
annihilates (creates) an electron in a spin-$\sigma$ Wannier state 
at the lattice site ${\bf R}_i$.
The second part stands for the interaction between the impurity spin
${\bf S}_0$ and the spins of the conduction electrons represented by 
${\mbox{\boldmath${\sigma}\!$\unboldmath}\,}_0$.
Throughout this paper a spin-$\frac{1}{2}$ impurity will be considered.
We assume 
a $k$-independent 
coupling constant $J$. Therefore the coupling is effectively local,
involving only 
electrons 
at the impurity
site ${\bf R}_0$. 
${\cal H}$ represents the (antiferromagnetic) Kondo
model if $J<0$, and the ferromagnetic Kondo model for $J>0$.

\subsection{Exact solution for an
empty conduction band}
\label{subsec:Theory_ngl0}

Appropriate Fourier transformations of the equations of motion for the 
Green's functions 
$\Geup{ij}(E) = {\GFjup{\ceup{i}}}_E$ and 
$\GAup{0ij}(E) = {\GFjup{\eSz{0} \ceup{i}}}_E$ 
lead to the following
closed 
system of equations:
\begin{align} 
\Geup{jj}
&=
G_{00}^{(0)}
-
G_{j0}^{(0)}
\frac{3}{2} J \,  
\GAup{00j}
\label{eq:G_jj_von3}
\\
\Geup{0j}
&=
G_{00}^{(0)}
-
G_{00}^{(0)}
\frac{3}{2} J \,  
\GAup{00j}
\label{eq:G_0j_von3}
\\
\GAup{00j} 
&=
- 
G_{00}^{(0)}
\frac{J}{8}  
\Geup{0j}
+ 
G_{00}^{(0)}
\frac{J}{2}  
\GAup{00j} \;\;.
\label{eq:GA_00j_von3}
\end{align}
The free Green's functions are given by $G_{ij}^{(0)}(E)=\frac{1}{N}
\sum\limits_{{\bf k}} 
\frac{e^{i{\bf k}({\bf R}_i-{\bf R}_j)}}{E-\epsilon({\bf k})
+i0^+}$, with 
$N$ the 
number of lattice sites.
Solving 
(\ref{eq:G_jj_von3}) -- (\ref{eq:GA_00j_von3})
yields the 
solution:
\begin{equation}
G_{jj\uparrow} = G_{00}^{(0)} -  
\frac
{ 
G_{00}^{(0)} \frac{3}{16} J^2 
{\big( G_{j0}^{(0)} \big)}^2 
} 
{ \big(G_{00}^{(0)}\big)^2 \frac{3}{16} J^2 + 
{G_{00}^{(0)}} \frac{1}{2} J - 1 
}\;\;. 
\label{eq:Ge_jj_Ergebnis}
\end{equation}
With (\ref{eq:Ge_jj_Ergebnis}), the 
LDOS $\rho_{j\uparrow}$ 
of any given site ${\bf R}_j$ can be calculated:
$\rho_{j\uparrow}(E)=-\frac{1}{\pi} {\rm Im} \Geup{jj}(E)$.  
The spin dependence of 
$\rho_{j\uparrow}$ 
is purely formal.

For our calculations we have chosen the tight-binding dispersion with
nearest-neighbour hopping for the s.c. lattice:
\begin{eqnarray}
\epsilon({\bf k})=-2t(\cos k_x a + \cos k_y a + \cos k_z a) 
\label{eq:freieDispersion}
\end{eqnarray}
where $t$ is the hopping integral and $a$ the
lattice constant. (\ref{eq:freieDispersion})
implies a symmetric Bloch density of states (BDOS) and
${\rm Im} G_{i0}^{(0)}(E)=\pm {\rm Im} G_{i0}^{(0)}(-E)$, which
result in 
a mirror symmetry of the LDOS
when changing 
from the antiferromagnetic to the ferromagnetic Kondo model:
$\rho_{i\uparrow}^{(J)}(E)=\rho_{i\uparrow}^{(-J)}(-E)$ at any site
${\bf R}_i$.

\subsection{Approximation scheme for finite band occupation}
\label{sec:Appr_scheme}

For $\Geup{ij}$ 
Eqs.~(\ref{eq:G_jj_von3}) and (\ref{eq:G_0j_von3}) remain 
correct. 
For $\GAup{0ij}$
two new
Green's functions 
come into play, to which
Nagaoka's de{\nolinebreak}coupling scheme is applied
\cite{Nag65}.
Consequently, Eq.~(\ref{eq:GA_00j_von3}) is replaced by 
\begin{align} 
&\GAup{00j}(E) 
\approx
\nonumber\\ 
&\approx \Big(\! - 
G_{00}^{(0)}(E)
\frac{J}{8}  
+ J
\la \, S_0^z \cekup{0} \ceupT{0}(E) \, \ra \Big) \,
\Geup{0j}(E)
\nonumber
\\
&+ 
\Big(
G_{00}^{(0)}(E)
\frac{J}{2}  
- J 
\la \, \cekup{0} \ceupT{0}(E) \, \ra \Big) \,  
\GAup{00j}(E) \;.
\label{eq:GAeej_gl_sumkKorrT}
\end{align} 
We 
have defined special energy-dependent annihilation operators  
\begin{eqnarray}
\tilde{c}_{0\sigma}^{}(E)
\stackrel{\scrscr \rm def.}{=}
\frac{1}{\sqrt{N}}
\sum_{{\bf k}}
\frac{e^{i{\bf k}{\bf R}_0}}
     {E-\epsilon({\bf k})+i0^+}
\, c_{{\bf k}\sigma} \;
\label{eq:mod_operator}
\end{eqnarray}
to avoid multiple $k$-space summations within the 
self-consistency cycle of this approximation scheme.
The remaining problem is to determine the correlation functions of
Eq.~(\ref{eq:GAeej_gl_sumkKorrT}).
They are complex quantities, whose real and imaginary parts must be 
determined 
separately via the spectral theorem with the help of 
appropriately defined operators and Green's functions. For example:
\begin{align} 
&{\rm Re} \la \, \cekup{0} \ceupT{0}(E) \, \ra
=
\la \, \cekup{0} \ceupR{0}(E) \, \ra
\label{eq:Re_R}
\\
\mbox{with}&\;\; \stackrel{{}\, \scrscr R}{c}_{0\sigma}^{}(E)
\stackrel{\scrscr \rm def.}{=}
\frac{1}{\sqrt{N}}
\sum_{{\bf k}}
{\cal P}
\frac{e^{i{\bf k}{\bf R}_0}}
     {E-\epsilon({\bf k})}
\,c_{{\bf k}\sigma} \;,
\end{align}
${\cal P}$ meaning ``principal part of''.
One needs 
\begin{equation}
\GeupR{00}(E,E^\prime) = \las \ceupR{0}(E) \,;\, \ceup{0} \ras_{E^\prime}
\end{equation} 
with $E$ a parameter and $E^\prime$ the usual energy variable of the
Green's function
to determine
$\la \, \cekup{0} \ceupR{0}(E) \, \ra$ 
via the spectral theorem
\begin{eqnarray} 
\la \, \cekup{0} \ceupR{0}(E) \, \ra 
=
- \frac{1}{\pi}
\int\limits_{-\infty}^{+\infty}
\frac{{\rm Im}\GeupR{00}(E,E^\prime)}
     {e^{\beta(E^\prime-\mu)}+1}
dE^\prime \;.
\label{eq:ccR_Spektralth}
\end{eqnarray} 
Comparing the respective equations of motion for $\GeupR{00}(E,E^\prime)$ and
$\Geup{00}(E)$ one can show that
$\GeupR{00}(E,E^\prime)$ is 
fully 
determined by the ``simple''
one-electron Green's function through 
\begin{eqnarray}
\GeupR{00}(E,E^\prime)
=
\;
\frac{\GOR{00}(E,E^\prime)}
     {G_{00}^{(0)}(E^\prime)}
\,\Geup{00}(E^\prime) \;,
\label{eq:GeeR_Gee}
\end{eqnarray}
with
\begin{align}
&\GOR{00}(E,E^\prime)
\!
=
\!
\frac{1}{N}
\sum_{\bf k}
\!
\frac{1}{E^\prime-\epsilon({\bf k})+i0^+}
{\cal P}
\frac{1}{E-\epsilon({\bf k})} \,.
\end{align}
Thus ${\rm Re} \la \, \cekup{0} \ceupT{0}(E) \, \ra$ 
and, analogously, ${\rm Im} \la \, \cekup{0} \ceupT{0}(E) \, \ra$ are 
fully 
determined by $\Geup{00}$.

To determine the real part of the other correlation
function, ${\rm Re}\la \, S_0^z \cekup{0}
\ceupT{0}(E) \, \ra$, one needs 
the Green's function
${\GAupR{000}(E,E^\prime)=\las S_0^z \ceupR{0}(E) \,;\, \cekup{0} 
\ras_{E^\prime}}$. 
We assume the same proportionality (\ref{eq:GeeR_Gee})
that holds between $\GeupR{00}(E,E^\prime)$ and $\Geup{00}(E^\prime)$
for 
the $\Gamma$-Green's functions:
\begin{eqnarray}
\GAupR{000}(E,E^\prime)
\approx
\;
\frac{\GOR{00}(E,E^\prime)}
     {G_{00}^{(0)}(E^\prime)}
\,\GAup{000}(E^\prime) \;\;.
\label{eq:GAeeR_GAee}
\end{eqnarray}
Note that Eq.~(\ref{eq:GAeeR_GAee})
is 
required to
work 
only within the spectral-theorem integration over $E^\prime$
to determine ${\rm Re} \la \, S_0^z \cekup{0} \ceupT{0}(E) \, \ra$.
Within that integration great contributions 
arise for $E^\prime\approx E$, for which (\ref{eq:GAeeR_GAee})
can be shown to 
be reasonably justified.
Analogous considerations 
can be made 
for ${\rm Im} \la \, S_0^z \cekup{0} \ceupT{0}(E) \, \ra$.

Hence, 
one has 
a closed 
system of equations consisting of 
(\ref{eq:G_jj_von3}), (\ref{eq:G_0j_von3}),
(\ref{eq:GAeej_gl_sumkKorrT}), (\ref{eq:GeeR_Gee}),
(\ref{eq:GAeeR_GAee}), the analogues to (\ref{eq:GeeR_Gee}) and
(\ref{eq:GAeeR_GAee}) 
for determining ${\rm Im} \la \, \cekup{0} \ceupT{0}(E) \, \ra$ and
${\rm Im} \la \, S_0^z \cekup{0} \ceupT{0}(E) \, \ra$,
and the spectral theorems (such as (\ref{eq:ccR_Spektralth})) for the real and
imaginary parts of  
the correlation functions. 
The LDOS $\rho_{i\uparrow}(E)=-\frac{1}{\pi} {\rm Im}\Geup{ii}(E)$ can
be self-consistently determined for any lattice site ${\bf R}_{i}$. 
The spin dependence of 
$\rho_{j\uparrow}$ 
is purely formal.
The occupation number at ${\bf R}_{i}$ is gained by
integrating over the LDOS:
\begin{eqnarray}
\langle n_i \rangle= 2* \int\limits_{-\infty}^{+\infty} dE \,
\frac{\rho_{i\uparrow}(E)}{e^{\beta (E - \mu)} + 1} \;.
\end{eqnarray}

\section{Results and Discussion}
\subsection{LDOS in the limiting case of 
an empty conduction band}
\label{subsec:empty_band}

\begin{figure}[h]
\includegraphics[width=0.9\linewidth]{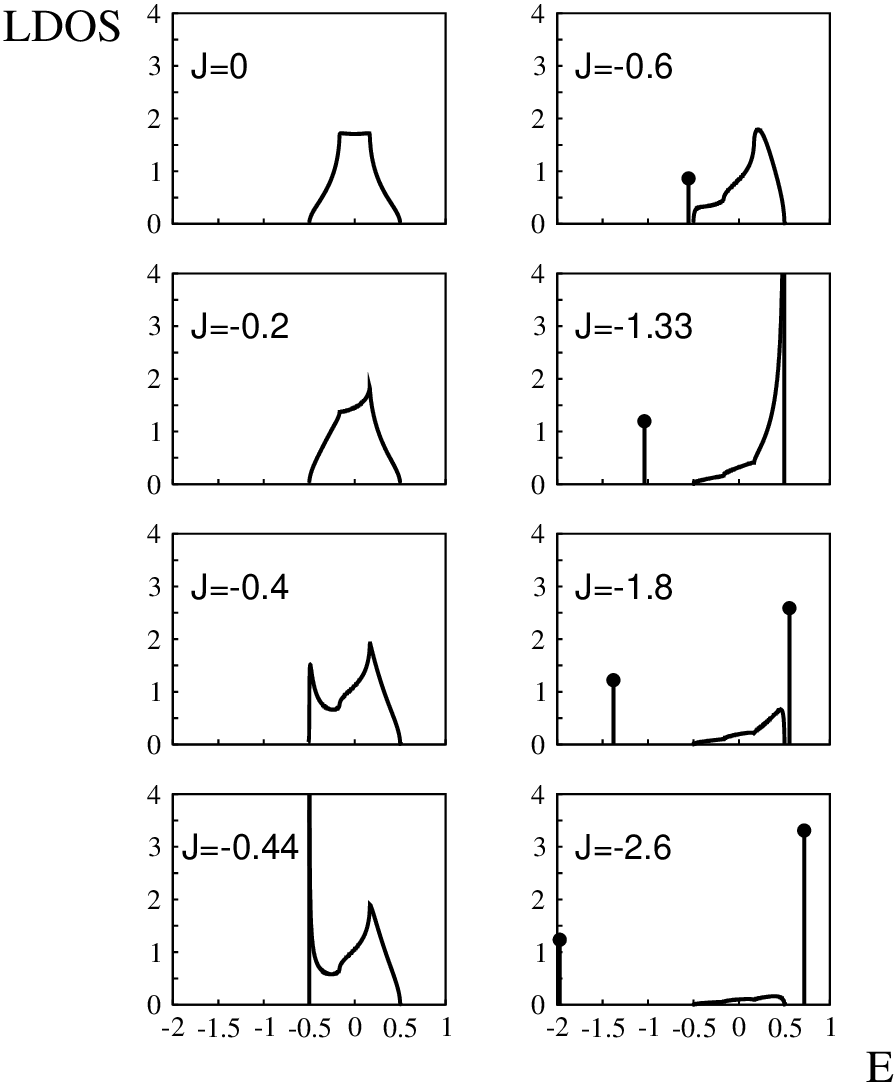}
\caption{LDOS of the antiferromagnetic Kondo model at the impurity site 
in the limiting case of an empty conduction band 
for various values of the coupling constant $J$.
$\delta$-peaks are represented by pins whose heights
are proportional to the 
corresponding spectral weights. 
The free Bloch dispersion is 
given by Eq.~(\ref{eq:freieDispersion}). 
The energy is 
measured in units of the bandwidth.}
\label{fig:n0r000Jn}
\end{figure}
\begin{figure}[h]
\includegraphics[width=0.9\linewidth]{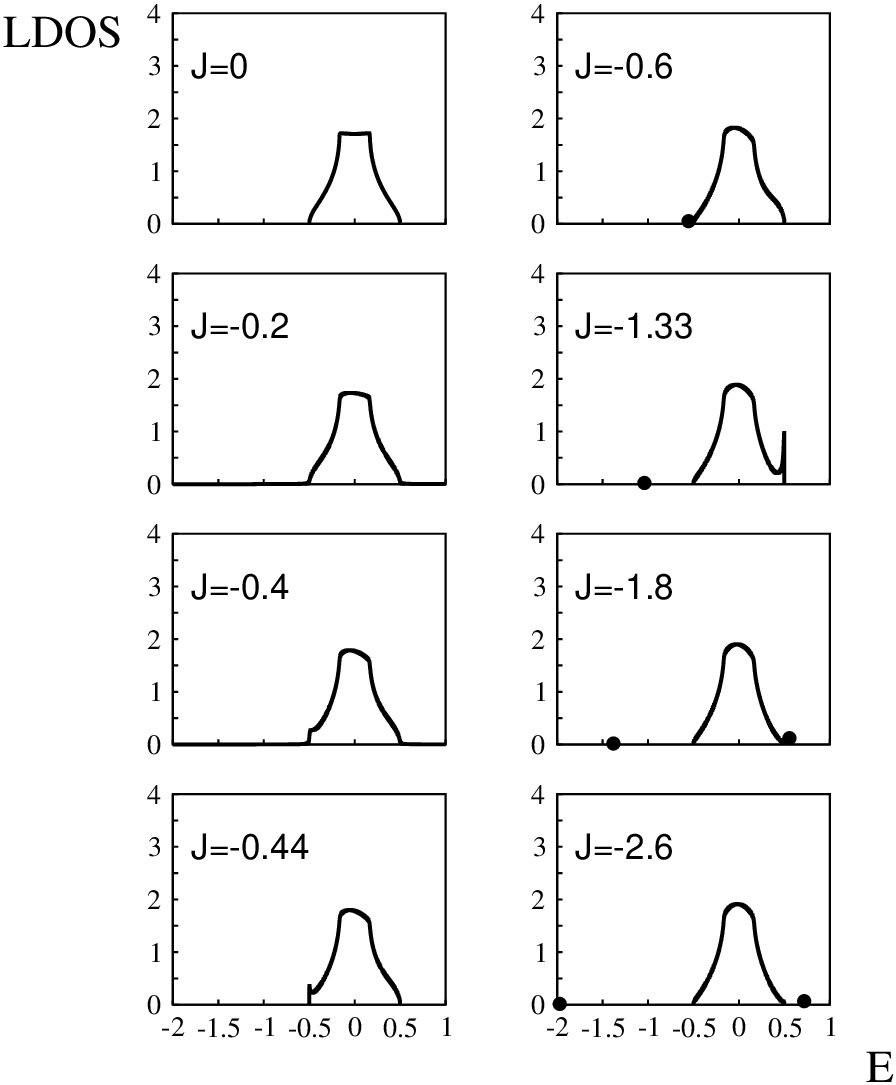}
\caption{LDOS of the antiferromagnetic Kondo model at a nearest-neighbour 
site of 
the impurity in the limiting case of an empty conduction
band. The rest as in Fig.~\ref{fig:n0r000Jn}.}
\label{fig:n0r100Jn}
\end{figure}
We first discuss 
the limiting case of an empty conduction band,
in which results 
are exact and 
allow rigorous interpretation, and which is similar in the 
antiferromagnetic and ferromagnetic Kondo models.
The LDOS 
represents the outcome of an inverse-photoemission 
experiment in which one test-electron is put into the 
empty conduction band.
It thus provides information on the impurity's effects 
on one-electron states. 
In Figs.~\ref{fig:n0r000Jn} 
and \ref{fig:n0r100Jn} the LDOS of the antiferromagnetic Kondo model at the impurity 
and the nearest-neighbour sites is shown 
for various coupling constants $J<0$.
The LDOS of the ferromagnetic Kondo model ($J>0$) is simply given by 
the mirror image 
of 
the 
LDOS of the antiferromagnetic Kondo model with the same 
$|J|$. 
This symmetry is a consequence of the 
Bloch dispersion
we have chosen 
(Eq.~(\ref{eq:freieDispersion})).
A general dispersion gives similar 
but not exactly symmetric results for the two models.

At $J=0$  the LDOS 
corresponds to the free Bloch density of states (BDOS).
Small couplings 
cause deformations of the LDOS,
most pronounced at the impurity site.
For $J<-0.44$ there appears a $\delta$-peak below, for 
$J<-1.33$ a second $\delta$-peak above the quasiparticle band.
The upper peak has greater spectral weight than the lower. 
At the nearest-neighbour site the weight of both peaks 
is considerably less than at the 
impurity site.
With increasing $|J|$ the peaks move away from the quasiparticle band.
At the impurity site their
weight 
increases,
while the weight of the quasiparticle band gradually vanishes.
At the neighbouring site the $\delta$-peaks lose weight (not
recognizable in Fig.~\ref{fig:n0r100Jn}) and the quasiparticle band 
approaches a certain asymptotic form.

This 
can be well understood 
from 
the strong-coupling limit (${J\to -\infty}$).
In that limit the singlet and triplet states with the  
electron 
completely localized at the impurity site are 
energy eigenstates.
The other one-electron states are extended 
band states,
deformed in the vicinity of and zero at the impurity site,
thereby meeting the condition of orthogonality to the 
fully localized states. 
Therefore, in the strong-coupling limit the 
LDOS at the impurity site
consists of two
$\delta$-peaks,
their positions tending towards $-\infty$
and $+\infty$, corresponding to 
the diverging energies of 
the bound singlet and triplet states.
The quasiparticle band 
has zero weight 
as there is no overlap of the band states to the impurity site.
At the other sites the 
LDOS vice versa consists of a quasiparticle band 
only. It is deformed 
because the probability of the electron to stay 
at the respective site is redistributed between and within the extended
states.

\begin{figure}[h]
\centerline{
\includegraphics[width=0.8\linewidth]{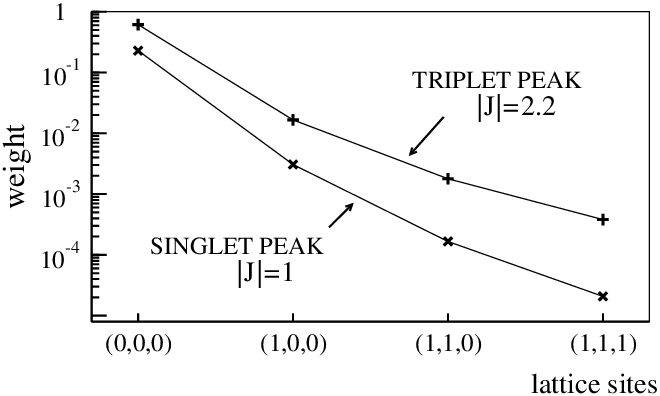}
}
\caption{Quasi-exponential decrease of the spectral weights of the singlet and triplet peaks with increasing distance 
  from the impurity site (0,0,0).
The weight of the singlet peak is
  shown for $|J|=1$, the weight of the triplet peak for $|J|=2.2$.
(Calculated for same Bloch dispersion 
as in Fig.~\ref{fig:n0r000Jn}.)}
\label{fig:gewS-1_0T-2_2}
\end{figure}
For finite $J$
the bound states 
have finite overlaps to the neighbouring sites of the impurity, 
which is reflected 
by quasiparticle $\delta$-peaks with small but finite 
weights.
Since the eigenstates with one electron can be classified as pure 
singlet or triplet states for all $J$, 
the lower and upper
quasiparticle $\delta$-peaks correspond to bound 
singlet 
and 
triplet
states, respectively. 
The greater weight of the upper peak (``triplet peak'')
results from the degeneracy of the triplet states.
The degree of localization 
of the bound states around the impurity site is very large
as can be seen when comparing the spectral weights of the singlet and
triplet peaks at the impurity 
and neighbouring sites (Fig.~\ref{fig:gewS-1_0T-2_2}).
There is a quasi-exponential decrease of weights/overlaps
with increasing distance from the impurity. 
The extended band states in case of finite $J$ have a finite overlap 
to the impurity site, 
which is reflected by the finite weight of the quasiparticle band 
in the impurity-site LDOS.

If considering singlet and triplet states separately,
the situation is 
similar to 
the quantum-mechanical 
problem of a
particle 
in a potential landscape with 
a well or barrier. 
For instance, the bound singlet state corresponds to a bound state 
in a potential well, 
which has an exponentially decreasing overlap 
into the adjacent region if the depth of the well is finite. 
In case of an infinitely deep well, 
corresponding to the strong-coupling limit,
each bound state is 
restricted to the well while
the extended states
have no overlap into the well region.

\begin{figure}
\centerline{
\includegraphics[width=0.8\linewidth]{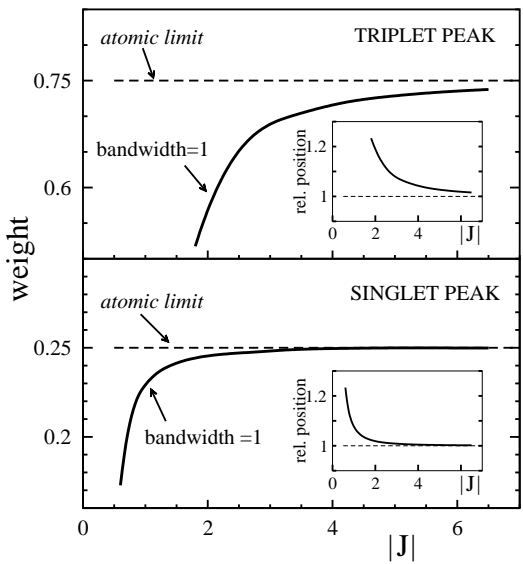}
}
\caption{Convergence of the weights and positions of the impurity-site 
singlet and triplet peaks toward the respective atomic-limit values.
The solid lines in the insets show 
the {\em relative positions} of the peaks 
(ratio of position in bandwidth=1-case and atomic limit).
(Calculated for same 
Bloch dispersion as in Fig.~\ref{fig:n0r000Jn}.)}
\label{fig:delAtLim}
\end{figure}
For large values of $J$ the effect of the hopping from and to the 
impurity site becomes negligible compared to the spin-spin coupling.
Therefore,
as far as the impurity site is concerned, 
a comparison with the atomic limit \cite{NoM84}
is sensible.
In case of zero band occupation the LDOS in the atomic limit consists of 
two quasiparticle $\delta$-peaks 
whose positions are at $+\frac{3}{4}J$ (singlet) and $-\frac{1}{4}J$ 
(triplet), having 
spectral weights of $\frac{1}{4}$ and $\frac{3}{4}$, respectively.
In Fig.~\ref{fig:delAtLim} it is shown how  
the weights and positions of the 
singlet and triplet peaks in the impurity-site LDOS 
converge for large $J$ towards the corresponding atomic-limit values, 
which confirms the 
emergence of fully localized (atomic) states in the strong-coupling 
limit. 

\subsection{LDOS in the antiferromagnetic Kondo model}
\label{sec:afmKM}

\begin{figure*}
\includegraphics[width=0.8\linewidth]{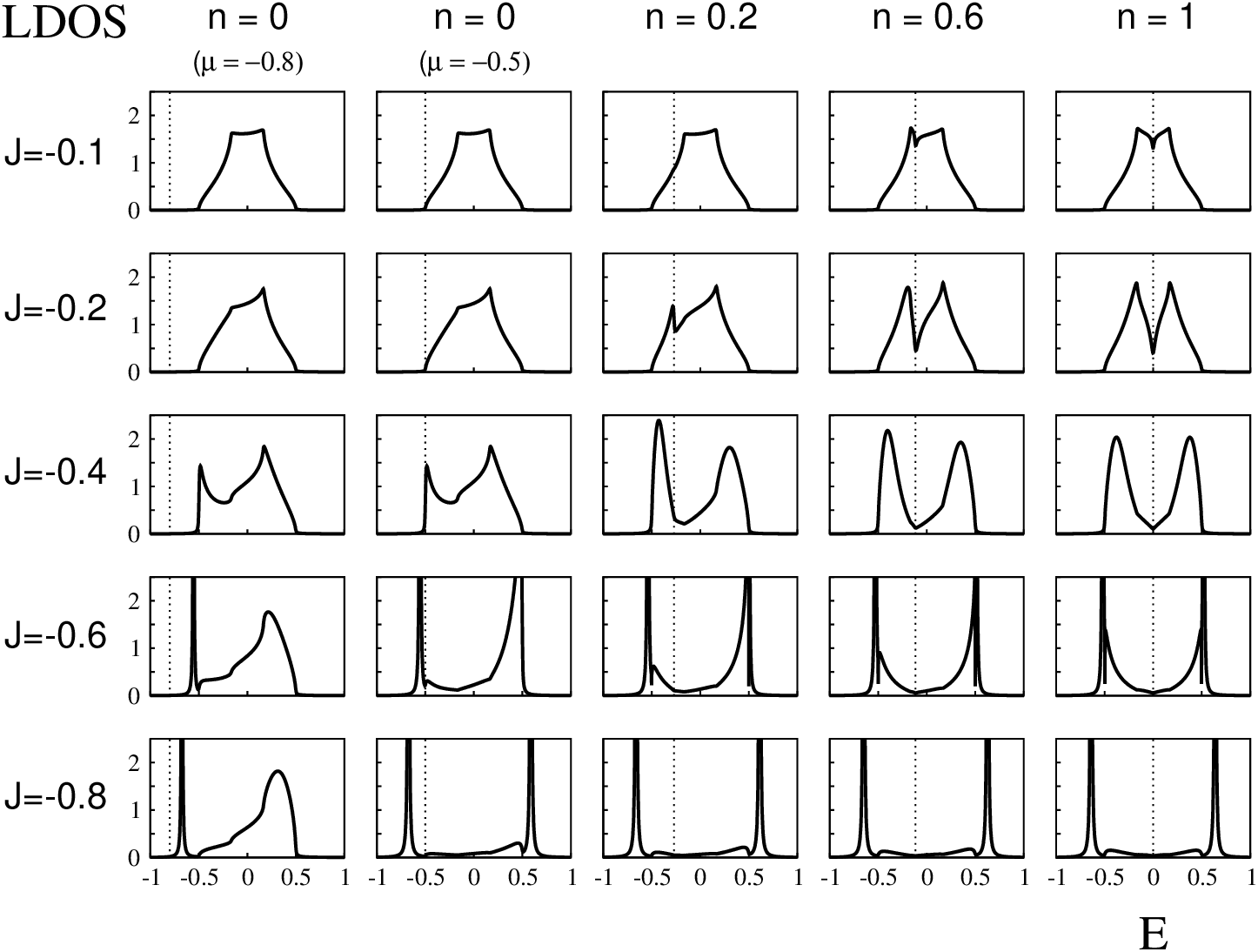}
\caption{LDOS of the antiferromagnetic Kondo model at the impurity site 
for various coupling constants $J$ and 
band occupations $n$. The position of the 
chemical potential $\mu$ is indicated by the thin dotted line.
The narrow resonances outside the quasiparticle band 
are quasiparticle $\delta$-peaks. 
(Calculation for 
the same Bloch dispersion as in 
$\mbox{Fig.~\ref{fig:n0r000Jn}}$.)}
\label{fig:rn0Jn}
\end{figure*}

In Fig.~\ref{fig:rn0Jn} our results for the Kondo-model LDOS at the
impurity site for different band occupations $n$ and the full range of
weak to strong couplings $J$ are presented.
At more than half filling ($n>1$) the LDOS is simply given by the
mirror image of the corresponding LDOS 
at a band occupation of $2-n$ (due to particle-hole symmetry of the
Kondo Hamiltonian (\ref{eq:H-op}) 
and the selection 
of a symmetric free BDOS).
For weak to intermediate couplings 
($|J|\le 0.4$) and not too small band
occupations ($n\ge 0.2$) the most 
prominent feature in the LDOS is an
antiresonance at the Fermi level,
which will be discussed in 
subsection 1.
In the strong-coupling regime ($|J|\ge 0.6$)
quasiparticle $\delta$-peaks 
are the dominant characteristic, 
which are discussed 
in 
subsection 2.

\subsubsection{Kondo antiresonance}
\label{sec:Kondo_antiresonance}

\begin{figure}[h]
\centerline{
\includegraphics[width=0.55\linewidth]{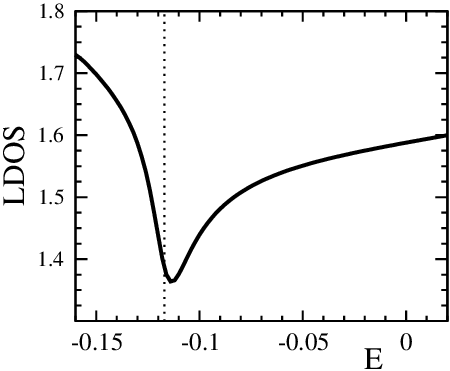}
}
\caption{Antiresonance in the impurity-site LDOS of the antiferromagnetic Kondo 
  model (${n=0.6, \,J=-0.1}$). The Fermi level is indicated by the thin
  dotted line. The shift of the
  antiresonance relative to the Fermi level is of the order of $10^{-3}$
in units of the bandwidth. 
(Same calculation as in  
$\mbox{Fig.~\ref{fig:rn0Jn}}$.)}
\label{fig:shiftanti1}
\end{figure}

For small couplings ($J= -0.1$) 
the LDOS 
exhibits a narrow, dip-like structure 
at the Fermi
level. When
increasing the coupling strength ($J= -0.2, -0.4$) 
it deepens and broadens, 
developing into a marked antiresonance. 
As shown in $\mbox{Fig.~\ref{fig:shiftanti1}}$, 
the antiresonance appears not exactly at the Fermi energy but is
slightly shifted (upward shift for $n<1$).
For strong couplings ($|J|\ge 0.6$) the LDOS at 
the Fermi level is practically zero. However, no
antiresonance in the narrower 
sense is left, instead the LDOS is dominated by quasiparticle peaks. 
In the following paragraphs we will discuss 
independent evidence for an antiresonance 
in the (antiferromagnetic) Kondo model. After that, the 
antiresonance will be related to
a structure recently observed in STS experiments.

In Ref.~\onlinecite{MeZ71} Mezei and Zawadowski considered 
the Kondo model for
different types of scattering.
They made explicit statements on the 
LDOS for the resonance energy, for which they could refer to a
definite value of the phase shift.
For $s$-type scattering they found 
complete suppression of the LDOS, while
non-$s$-type scattering
was shown to 
involve no changes at the resonance energy. As each 
of 
the
scattering types ($s,p,$ etc.) is connected with a 
special wave-vector
dependence in the coupling constant, it 
seems a priori unclear how to apply 
those
results 
to the Kondo model
with a wave-vector independent coupling-constant $J$ discussed in this paper. 

Evidence for an antiresonance in the LDOS of the Kondo model
is 
provided by 
a comparison with results for 
the
(single-impurity) Anderson model \cite{And61}.
As is well-known, the Anderson model 
in the so-called Kondo limit can be mapped on the
Kondo model by the unitary Schrieffer-Wolff transformation \cite{ScW66}.
A correspondence 
must be expected between results in the 
Kondo limit of the Anderson model 
and 
the weak-coupling limit of the Kondo model.
In the Anderson model
one does observe an antiresonance 
in the conduction-electron LDOS.
It is 
closely related to the appearance of the well-known Kondo resonance
(peak) in the impurity 
quasiparticle density of states 
(impurity QDOS). 
Sollie and Schlottmann calculated 
the conduction-electron LDOS and the impurity QDOS of the 
single-impurity Anderson model within the slave-boson approach \cite{SoS90}.
The conduction-electron LDOS shows the antiresonance
appearing at the same energy  
and with the same width as the Kondo resonance in the impurity QDOS, 
which reveals the intimate relation between both structures.
Both appear not exactly at the Fermi energy but 
slightly above,
similar to what we observe for the
antiresonance in the Kondo model.
More recently Schiller and Hershfield emphasized the direct correspondence 
between antiresonance and Kondo resonance (``Abrikosov-Suhl resonance'') 
in the Anderson model \cite{ScH00}. 

\begin{figure}[h]
\centerline{
\includegraphics[width=\linewidth]{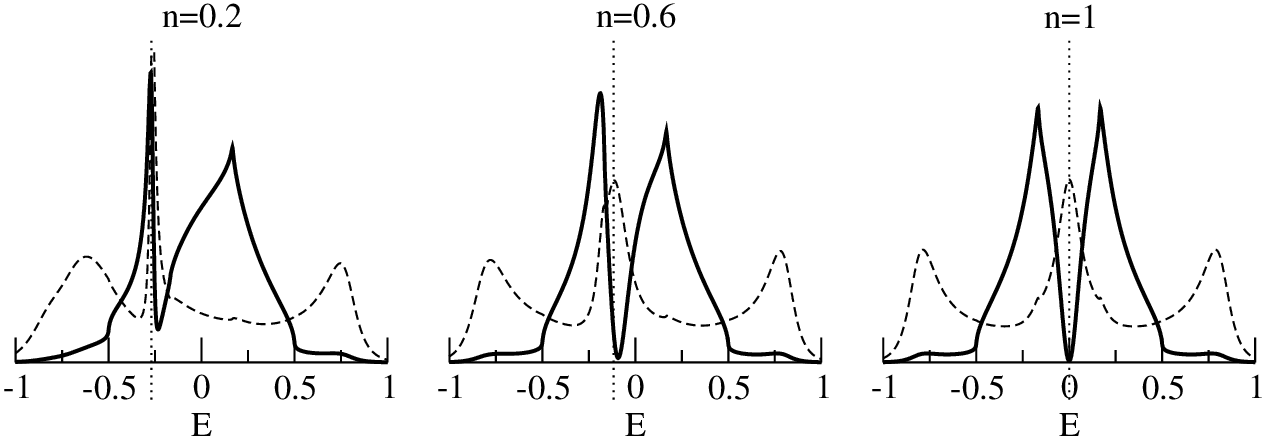}
}
\caption{Conduction-electron LDOS at impurity site (solid line) and
  impurity QDOS (broken line) in the Anderson model for different 
  band occupations $n$ calculated with the modified perturbation theory 
  (MPT). 
  The position of the Fermi level is indicated by the thin dotted line.
Parameters: $U=1$, $V=0.2$ (Bloch dispersion as in Fig.~\ref{fig:rn0Jn}).
}
\label{fig:siam_mpt}
\end{figure}

We have performed a calculation for the Anderson model within the modified
perturbation theory (MPT). The MPT is an extension of the conventional 
second-order perturbational scheme 
\cite{Hew,YoY70} to improve the
strong-coupling behaviour 
away from the symmetric point, which is the only point at
which the conventional perturbation theory also converges for large
Hubbard interaction $U$.\cite{ZlH83}
For details we refer to Ref.~\onlinecite{MWPN99}, 
where the MPT was successfully
applied to the single-impurity Anderson
model. It has also been used in the context of the dynamical mean-field theory
(DMFT) to examine the Hubbard
model \cite{PWN97} and the periodic Anderson model
\cite{MeN00Kon,MeN00Dyn}.
Our results for 
the single-impurity Anderson model for three different band occupations $n$
are shown in Fig.~\ref{fig:siam_mpt}.
The Kondo resonance in the impurity QDOS and 
the antiresonance in the conduction-electron LDOS are clearly to be seen 
at the Fermi level (for ${n=0.2, \, 0.6}$ slightly above).
With the parameters chosen,
an approximate correspondence to the Kondo model with $J=-0.16$ is to be 
expected according to
the Schrieffer-Wolff transformation. For each of the three band occupations 
the Anderson-model LDOS 
is indeed quite similar to the Kondo-model LDOS for $J=-0.2$ 
in Fig.~\ref{fig:rn0Jn}. 
The only basic difference 
is for the symmetric Anderson-model LDOS, where
the antiresonance 
goes right down to zero, 
whereas in the $n=1$ Kondo-model LDOS 
it does not. For the symmetric Anderson model 
one can show, using the fact that the electrons 
form a Fermi liquid, that the conduction-electron LDOS of the impurity site 
must be zero at the Fermi energy. Assuming this 
is conserved in the Schrieffer-Wolff transformation,
it is apparently a shortcoming of our approximation scheme 
that for small to intermediate couplings the symmetric Kondo-model 
LDOS is not fully suppressed at
the Fermi energy. 
Furthermore, one cannot expect our approximation scheme to
satisfy the 
subtle exponential dependence of the width of the antiresonance on the
coupling constant $J$ 
via 
the Kondo temperature
$T_K$ \cite{Hew}, which should occur 
in analogy to 
the Anderson model.
Nevertheless, the antiresonances 
coming out 
from the calculations
clearly indicate the correct 
antiresonance 
of the Kondo model. 
Since the antiresonance of the Kondo model
is closely connected with the 
one in the Anderson model 
and because of the intimate relation of the latter to the well-known
Kondo resonance, we 
use for 
both antiresonance structures the term ``Kondo antiresonance''.

Recently experimental evidence was found for the Kondo antiresonance in the 
conduction-electron LDOS of magnetic impurity adatoms on metal surfaces.
In scanning tunneling spectroscopy (STS) experiments 
the differential conductance $dI/dV$ 
was measured for the cases of Ce adatoms on a Ag(111) surface 
\cite{LSBD98}, for Co on Au(111) 
\cite{MCJCW98} and for Co on Cu(111) 
\cite{MLE00}. 
In each experiment a narrow antiresonance structure 
with a slight upward shift relative to the Fermi level
was observed in the $dI/dV$ spectrum. 
Following Ref.~\onlinecite{ScH00},
in case of weak tunneling, low enough
temperatures ($T\to 0$) and in an energy range in which the LDOS of the
STM tip is nearly constant, the differential conductance $dI/dV$ is in 
general proportional to what we shall call a ``mixed LDOS''.
Contributions to this ``mixed LDOS''
come from all electronic states (``tunneling channels'') of the substrate 
that are involved in the tunneling process.
The ``mixed LDOS'' is not just the sum over partial densities of states 
but also contains quantum-interference effects between the different 
tunneling channels. 

As for their measurements on 
Ce/Ag(111), Li et al.~assumed 
that they were ``locally
sensitive to the hybridized $sp$ conduction band'' \cite{LSBD98}.
This is supported by Schiller and Hershfield \cite{ScH00}, who achieved 
optimal theoretical agreement with the measured antiresonance curve
on the assumption that alternative tunneling 
into the impurity (Ce) $f$-orbital was zero.
Accordingly, the pure conduction-electron LDOS was probed,
which allows to directly identify the antiresonance in the $dI/dV$ 
spectrum with the Kondo antiresonance in the conduction-electron LDOS.

By contrast, in the case of Co/Au(111) the involvement of 
the tunneling channels and 
the interpretation of the $dI/dV$ spectrum are a matter of 
controversy.
Madhavan et al.~assumed significant contributions from 
tunneling both into the conduction band and the impurity $d$-orbital of 
the Co adatom, attributing an asymmetry of the antiresonance 
to quantum-interference effects between the $d$-orbital and the conduction 
electron channels \cite{MCJCW98}.
This was confirmed by an excellent theoretical fit of the antiresonance.
However, \'Ujs\'aghy et al.~assumed 
tunneling into the conduction band only and achieved 
an equally good theoretical fit \cite{UKSZ00}.
The ambiguity is already inherent in
Fano's theory \cite{Fan61},
which in Refs.~\onlinecite{MCJCW98} and \onlinecite{UKSZ00} 
was generalized to the interacting case.
For non-interacting electrons one can show that 
the asymmetric shape
of a ``Fano resonance'' depends,
first, on the ratio of 
tunneling into the impurity orbital and tunneling into the conduction 
band, and secondly, 
on the symmetry/asymmetry 
of the free conduction-electron BDOS and the Fermi-level position.
Each of these factors and 
combinations of the two can lead to the same 
line shape.
This seems to 
hold in the present (interacting) case, too.
We conclude that 
in the case of Co/Au(111) the $dI/dV$ spectrum possibly represents a
direct measure of the conduction-electron LDOS. 
If so, the asymmetric antiresonance (Fano resonance) 
directly represents a Kondo antiresonance. 
If not, a Kondo antiresonance 
may still be considered as
the main underlying structure 
apart from impurity-orbital contributions and quantum-interference 
effects.

\subsubsection{Quasiparticle $\delta$-peaks}
\label{subsubsec:afm_KM_delta_peaks}

\begin{figure*}[t]
\centerline{
\includegraphics[width=0.7\linewidth]{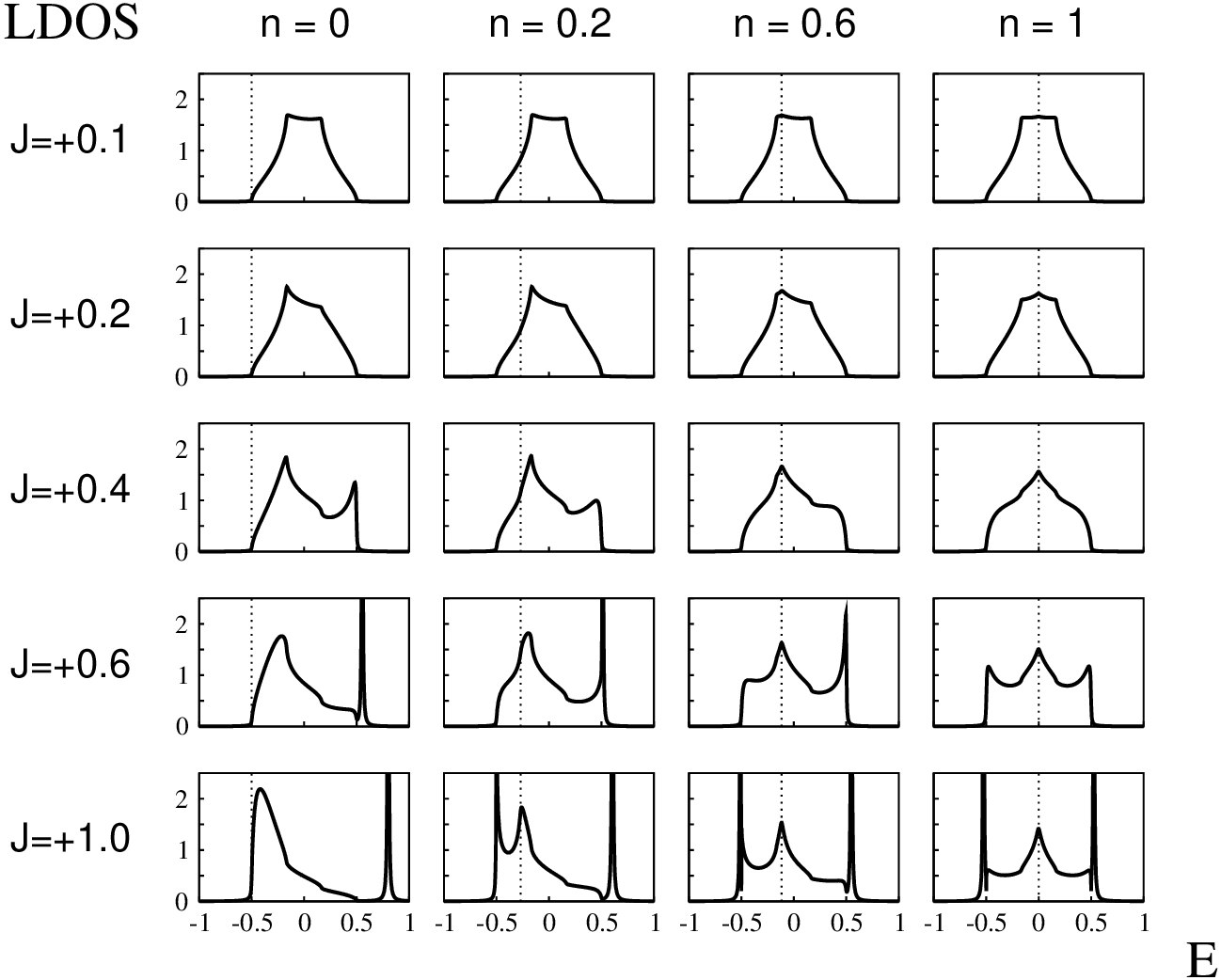}
}
\caption{LDOS of the ferromagnetic Kondo model at the impurity site 
for various coupling constants $J$ and 
band occupations $n$. The position of the 
chemical potential is indicated by the thin dotted line.
The narrow resonances outside the quasiparticle band 
are quasiparticle $\delta$-peaks. 
(Calculation for 
the same Bloch dispersion as in 
$\mbox{Fig.~\ref{fig:n0r000Jn}}$.)}
\label{fig:rn0Jp}
\end{figure*}

In the strong-coupling regime ($|J|\ge0.6$) the LDOS of the Kondo 
model is dominated by quasiparticle $\delta$-peaks. 
In Fig.~\ref{fig:rn0Jn} they appear as narrow resonances.
For the limiting case of an empty conduction band
(Sec.~\ref{subsec:empty_band}) we have 
discussed the emergence of a singlet and a triplet peak, corresponding to 
the excitation of bound one-electron singlet and triplet eigenstates through
inverse photoemission.
With increasing $J$ first the singlet peak emerges from the 
quasiparticle band.
In Fig.~\ref{fig:rn0Jn} this can be seen 
again in the first column on the left, which represents 
the limiting case of an empty conduction band.
If the singlet peak is present, 
one has to distinguish 
another case of zero band occupation
$n=0$ (second column in Fig.~\ref{fig:rn0Jn}), 
characterized by the 
chemical potential $\mu$ lying between the singlet peak and the 
quasiparticle band.
In this case there is exactly one electron in the system, occupying the 
bound singlet state, 
{\it before} the excitation in inverse or direct photoemission.
With this electron the band occupation $n$ (i.e.
the average number of
electrons per lattice site)
is still zero, but the physical situation differs from 
``the limiting case of an empty conduction band''.
This is reflected by the drastic differences between the corresponding
LDOS's in Fig.~\ref{fig:rn0Jn}.

For certain values of the coupling constant (cf. $J=-0.6$) the emergence 
of a second quasiparticle peak can be triggered by increasing 
the band occupation. The generic situation of the strong-coupling regime
is the presence of two quasiparticle peaks.
In the case of half filling they are positioned symmetrically above and 
below the quasiparticle band, whose weight gradually
vanishes with increasing coupling strength (cf. $J=-0.6, -0.8$).
The LDOS for large $J$ approaches the
LDOS of the atomic limit, which for $n= 1$ 
consists of two symmetric singlet peaks \cite{NoM84}, 
the lower one
observable in direct, the upper in inverse photoemission. 
The quasiparticle peaks 
of the strong-coupling regime clearly
correspond to the singlet peaks of the atomic limit. Nevertheless,
in the present case of finite hopping,  
admixtures of triplet coupling at the impurity site are to be
expected, which only disappear in the limit $J\to \infty$.
Note that 
these 
would not contradict the overall singlet nature of the
Kondo-model ground state.
In Ref.~\onlinecite{SoS90} similar quasiparticle peaks close to the band edges
were observed in the conduction-electron LDOS of the single-impurity
Anderson model.

\subsection{LDOS in the ferromagnetic Kondo model}
\label{sec:fmKM}

\begin{figure}[t]
\centerline{
\includegraphics[width=0.55\linewidth]{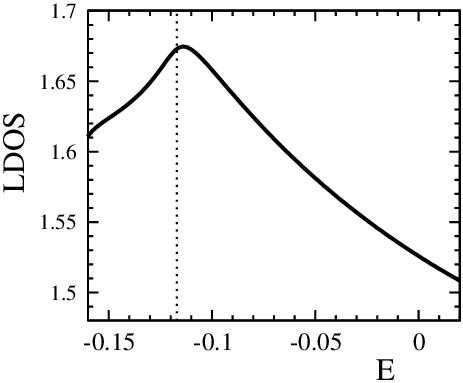}
}
\caption{Peak structure in the impurity-site LDOS of the ferromagnetic Kondo model 
(${n=0.6, \, J=+0.2}$). The Fermi level is indicated by 
the thin dotted line. 
The shift of the peak structure relative to the Fermi level is of the
order of $10^{-3}$ in units of the bandwidth.
(Same calculation as in $\mbox{Fig.~\ref{fig:rn0Jp}}$.)}
\label{fig:shiftreson}
\end{figure}

Figure \ref{fig:rn0Jp} shows our results for the impurity-site LDOS in the 
ferromagnetic Kondo model
for low to strong couplings $J$ and different band occupations $n$.
Due to particle-hole symmetry, it is sufficient to consider band occupations 
up to half filling. 
The most remarkable 
feature is a peak structure at the position of 
the Fermi level. 
Note that the peak structure is not a true resonance --- in the sense
that the LDOS 
is not really enhanced at the maximum in comparison to the free
BDOS. This agrees  
with what one should expect 
because of the vanishing phase shift at the Fermi
level in the ferromagnetic Kondo model. 
Instead, compared to the BDOS the LDOS 
is diminished 
in the vicinity of the maximum position.
The peak structure 
is slightly
shifted 
relative to the Fermi level 
(Fig.~\ref{fig:shiftreson}) similar to the Kondo antiresonance of the
antiferromagnetic Kondo model.

If the peak structure at the Fermi level is a true feature of the ferromagnetic Kondo model,
it should 
in principle 
be detectable in 
an STS experiment similar to the Kondo 
antiresonance
(cf. Sec.~\ref{sec:Kondo_antiresonance}). 
In an STS measurement the differential conductance $dI/dV$ 
represents a direct measure of the conduction-electron LDOS on condition
that tunneling occurs only into the conduction band. The peak structure in
the LDOS of the ferromagnetic Kondo model should thus be observable as a seeming
enhancement in the differential conductance near the Fermi level. 
However, the preparation of adatoms with {\it ferromagnetic} exchange 
($J>0$)
is required. Gd adatoms might be suitable candidates since in bulk Gd 
the intraatomic exchange is 
ferromagnetic, which can be inferred from an excess $T=0$ magnetic
moment due to induced ``ferromagnetic'' spin-polarisation of $5d-6s$
conduction electrons \cite{RCMM75}.

As in the antiferromagnetic Kondo model, for sufficiently large couplings $J$ 
quasiparticle $\delta$-peaks appear, 
whose emergence can partly be 
triggered by increasing the band occupation (cf.~$J=0.1$). 
Again, the generic situation 
in the strong-coupling regime is the presence of two quasiparticle
$\delta$-peaks.
In case of half filling they are located  
symmetrically below and above the quasiparticle band, whose weight
gradually vanishes when increasing the coupling strength.
Similar to the antiferromagnetic Kondo model and the limiting case of an empty
conduction band, the LDOS in the strong-coupling regime should approach 
the LDOS of the atomic limit. 
The two symmetric quasiparticle $\delta$-peaks therefore correspond to the 
two triplet peaks in the atomic limit \cite{NoM84} 
of the ferromagnetic Kondo model for $n=1$.
Nevertheless, in the case of finite hopping, admixtures
of singlet character are to be expected, which only disappear in the limit
$J\to \infty$.
The assumption of singlet admixtures is supported by 
the observation 
that for $J=1.0$ the upper 
peak in the ${n=1}$-LDOS 
continuously 
evolves from the pure singlet peak of the ${n=0}$-LDOS 
(limiting case of an empty conduction band)
when changing the band occupation from 0 to 1.

\subsection{Occupation numbers}
\label{sec:occup_numbers}
\begin{figure}[h]
\centerline{
\includegraphics[width=0.8\linewidth]{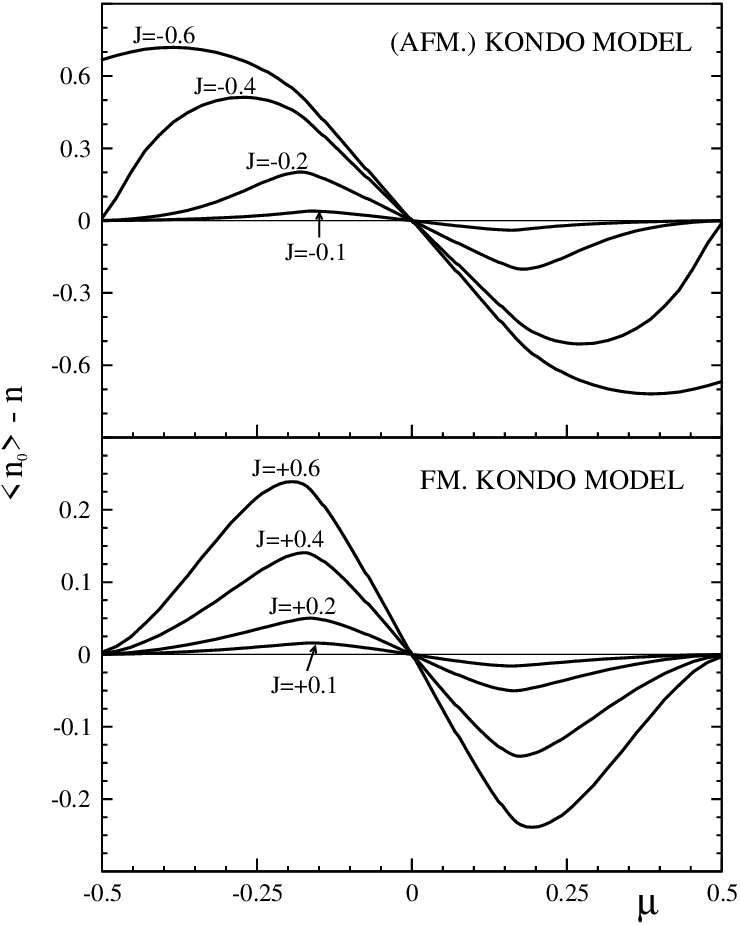}
}
\caption{Difference between the occupation number at the impurity site 
and the band occupation, ${\langle n_0 \rangle-n}$, in dependence on 
the Fermi level $\mu$ for different values of the coupling constant $J$. 
$\mu=-0.5$ corresponds to zero band-occupation, 
$\mu=0$ to half filling and ${\mu=+0.5}$ to a completely filled band. 
(Calculated for same 
Bloch dispersion as in Fig.~\ref{fig:n0r000Jn}.)}
\label{fig:dn0n_n}
\end{figure}
In Fig.~\ref{fig:dn0n_n} the effect of the magnetic impurity
on the occupation number $\langle n_0 \rangle$ at the impurity site  
is demonstrated.
Depending on the band occupation $n$, the impurity can effectively
attract or repel the conduction electrons both in the antiferromagnetic and ferromagnetic Kondo
models. 
For a symmetric BDOS the dependence on 
the band occupation is exactly symmetric: 
below half filling 
the impurity effectively attracts the conduction electrons,
$\langle n_0 \rangle - n > 0$ for $-0.5<\mu<0$,
above half filling
it repels them,
$\langle n_0 \rangle - n < 0$ for $0<\mu<0.5$.

The physical reason for this behaviour is clear and should analogously
hold
for a non-symmetric BDOS.
One of the possible couplings of 
an electron to the
impurity spin
(in the antiferromagnetic Kondo model the singlet, in the ferromagnetic Kondo model the triplet) 
decreases the energy of 
the system, making it favourable for the electron to stay at the
impurity site. 
Thus, 
$\langle n_0 \rangle$ is increased 
compared to the band occupation $n$
in the case of less than half filling.
However, it is unfavourable for a second electron to 
be at the impurity site 
(since it would break the favourable 
coupling of the 
other electron), which leads to an occupation number 
less than the band occupation in the case of more than half
filling. In a generalized form this argument implies 
a critical band occupation above which the occupation number 
at the impurity site is less than 
in the rest of the lattice, corresponding to 
a repulsive effect of the impurity on the conduction electrons.
Below we discuss 
results of Refs.~\onlinecite{EvG68} and \onlinecite{SZH89} which are 
seemingly contradictory to this. 

\begin{figure}[h]
\centerline{
\includegraphics[width=0.8\linewidth]{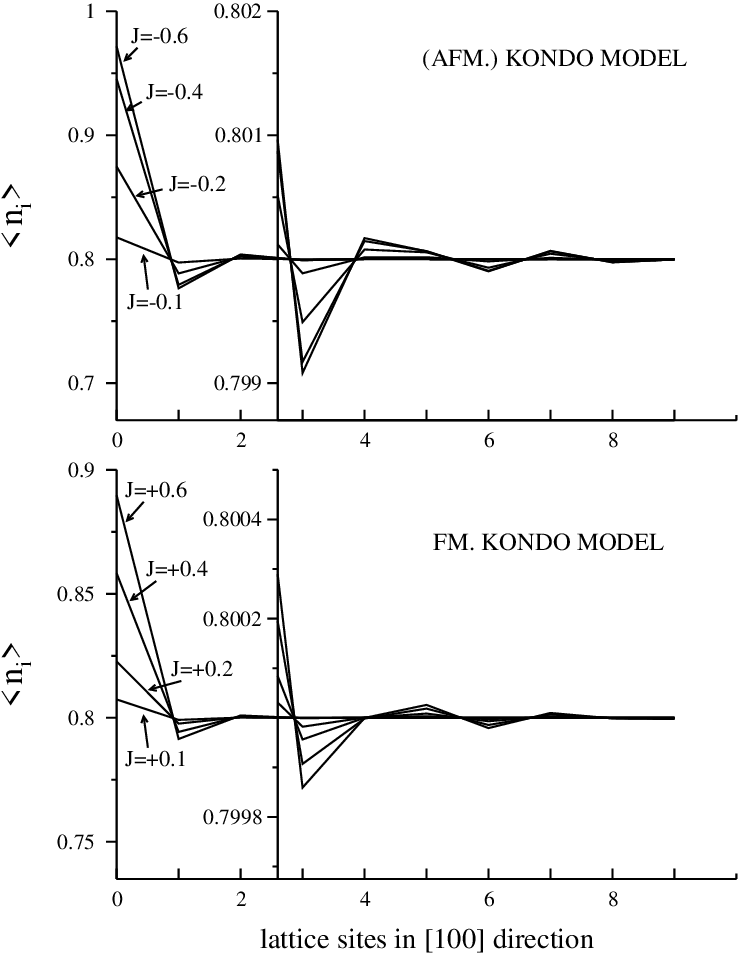}
}
\caption{Oscillations in the occupation number $\langle n_i \rangle$ for 
successive lattice sites in the [100] direction at a band occupation 
of $n=0.8$ for different coupling constants $J$. 
(Calculation for s.c. lattice with 
same Bloch dispersion as in Fig.~\ref{fig:n0r000Jn}.)}
\label{fig:oszill}
\end{figure}
The effect of the magnetic impurity on the occupation numbers $\langle
n_i \rangle$ 
at the surrounding lattice sites 
is demonstrated in Fig.~\ref{fig:oszill}. 
Oscillations around the value of the band occupation
which decrease in amplitude with growing distance from the impurity
are to be seen. 
The oscillations in the occupation number 
reflect the Friedel oscillations in the charge density,
which are 
the typical reaction 
of an electron system to 
an impurity potential.
Everts and Ganguly derived an expression for 
the charge density in the Kondo model (Ref.~\onlinecite{EvG68}).
\v{S}ok\v{c}evi\'c, Zlati\'c and Horvati\'c calculated 
charge-density oscillations 
in the Anderson model (Ref.~\onlinecite{SZH89}). 
A quantitative comparison between
charge-density and
occupation-number results is 
only partly possible.
The occupation number
is the number of electrons in the 
Wannier states at a given lattice site,
whereas the charge density corresponds to 
the number of electrons per unit volume for a given point of space. 
In contrast to the occupation number,
the charge density at and near 
a site depends also on 
electrons in Wannier states 
of other sites because in real space Wannier wave-functions are not 
confined to ``their'' sites.
Although there is no rigorous proportionality between occupation number
and charge density, one should, nevertheless, expect correlations between 
the two 
as far as general features such as 
oscillations or pronounced enhancements/reductions are concerned. 

In view of such general correlations it may seem
unclear why in 
in the cited references
the charge density at the impurity
is {\it always enhanced} (even divergent) 
independent of the band occupation or $k_F$, $\epsilon_F$ etc.
It seemingly contradicts our above argument.
However, the result of permanent enhancement of the charge density at the
impurity site is
non-generic and 
a consequence of 
integrations 
without cut-off.
Integrations were performed over plane waves 
up to arbitrarily large wave vectors.
Thus 
an infinite number of bands is implied
with each one-electron state coupled to the impurity spin with practically 
the same coupling strength.
There is no critical band occupation in this effective infinite-band
model because
an arbitrarily large number of electrons 
can stay at the impurity being coupled to its 
spin in a favourable way.

\section{Conclusions}
\label{sec:conclusions}

We have provided 
results
for the 
local density of states (LDOS) in the antiferromagnetic and
ferromagnetic Kondo models, which has not been calculated before with
its full energy dependence.
Although 
calculations were performed 
within a simple approximation scheme, 
results prove at least in qualitative terms trustworthy,
allowing consistent interpretation and qualitative comparison with the
experiment.
There are clear quantitative limitations 
as indicated for example by the insufficient depth of the Kondo
antiresonance.  
Nevertheless, we would expect 
confirmation of the 
general 
features by quantitatively reliable methods like NRG or QMC.
 
In summary, for small to intermediate couplings there
is an 
antiresonance in the 
LDOS of the antiferromagnetic Kondo model (Kondo antiresonance),   
which has an analogue in the conduction-electron LDOS of the
Anderson model. The Kondo antiresonance can be identified with antiresonance
structures observed in STS spectra of single magnetic adatoms on the
assumption that tunneling occured mainly into the conduction band.
In the ferromagnetic Kondo model a peak structure 
appears close to the
Fermi level. It should be observable in appropriate STS experiments on 
adatoms with ferromagnetic intraatomic exchange (possibly Gd adatoms).
For strong couplings the LDOS in both models is dominated by
quasiparticle $\delta$-peaks, 
which can be clearly related to the quasiparticle peaks of the exactly
solvable case of an empty conduction band
and the atomic limit. Finally, we have given arguments 
for the impurity's attractive or repulsive effect on conduction electrons 
at the impurity site 
dependent on the band occupation and for 
seemingly contradictory results 
gained by integrating without band cut-off. 

\begin{acknowledgements}
We would like to thank W. Hofstetter for drawing our attention to 
Refs.~\onlinecite{LSBD98,MCJCW98,MLE00}. We acknowledge the support
of the DFG (Sonderforschungsbereich 290)
and the Volkswagen Foundation.  
One of us (D.~M.)\ gratefully acknowledges the support of the
Friedrich-Naumann Foundation. 
\end{acknowledgements}


\begin{thebibliography}{10}
\expandafter\ifx\csname bibnamefont\endcsname\relax
  \def\bibnamefont#1{#1}\fi
\expandafter\ifx\csname bibfnamefont\endcsname\relax
  \def\bibfnamefont#1{#1}\fi
\expandafter\ifx\csname url\endcsname\relax
  \def\url#1{\texttt{#1}}\fi
\expandafter\ifx\csname urlprefix\endcsname\relax\def\urlprefix{URL }\fi
\providecommand{\bibinfo}[2]{#2}
\providecommand{\eprint}[2][]{\url{#2}}

\bibitem{Zen51}
\bibinfo{author}{\bibfnamefont{C.}~\bibnamefont{Zener}},
  \bibinfo{journal}{Phys. Rev.} \textbf{\bibinfo{volume}{81}},
  \bibinfo{pages}{440} (\bibinfo{year}{1951}).

\bibitem{Kon64}
\bibinfo{author}{\bibfnamefont{J.}~\bibnamefont{Kondo}},
  \bibinfo{journal}{Prog. Theor. Phys} \textbf{\bibinfo{volume}{32}},
  \bibinfo{pages}{37} (\bibinfo{year}{1964}).

\bibitem{TsW83}
\bibinfo{author}{\bibfnamefont{A.~M.} \bibnamefont{Tsvelick}} \bibnamefont{and}
  \bibinfo{author}{\bibfnamefont{P.~B.} \bibnamefont{Wiegmann}},
  \bibinfo{journal}{Adv. Phys.} \textbf{\bibinfo{volume}{32}},
  \bibinfo{pages}{453} (\bibinfo{year}{1983}).

\bibitem{Hew}
\bibinfo{author}{\bibfnamefont{A.~C.} \bibnamefont{Hewson}},
  \emph{\bibinfo{title}{The Kondo Problem to Heavy Fermions}}
  (\bibinfo{publisher}{Cambridge University Press},
  \bibinfo{address}{Cambridge}, \bibinfo{year}{1997}).

\bibitem{Wil75}
\bibinfo{author}{\bibfnamefont{K.~G.} \bibnamefont{Wilson}},
  \bibinfo{journal}{Rev. Mod. Phys.} \textbf{\bibinfo{volume}{47}},
  \bibinfo{pages}{773} (\bibinfo{year}{1975}).

\bibitem{Andr80}
\bibinfo{author}{\bibfnamefont{N.}~\bibnamefont{Andrei}},
  \bibinfo{journal}{Phys. Rev. Lett.} \textbf{\bibinfo{volume}{45}},
  \bibinfo{pages}{379} (\bibinfo{year}{1980}).

\bibitem{Wie80}
\bibinfo{author}{\bibfnamefont{P.~B.} \bibnamefont{Wiegmann}},
  \bibinfo{journal}{Sov. Phys. JETP Lett.} \textbf{\bibinfo{volume}{392}},
  \bibinfo{pages}{1980} (\bibinfo{year}{1980}).

\bibitem{FrO86}
\bibinfo{author}{\bibfnamefont{H.~O.} \bibnamefont{Frota}} \bibnamefont{and}
  \bibinfo{author}{\bibfnamefont{L.~N.} \bibnamefont{Oliviera}},
  \bibinfo{journal}{Phys. Rev. B} \textbf{\bibinfo{volume}{33}},
  \bibinfo{pages}{7871} (\bibinfo{year}{1986}).


\bibitem{SSK89}
\bibinfo{author}{\bibfnamefont{O.}~\bibnamefont{Sakai}},
  \bibinfo{author}{\bibfnamefont{Y.}~\bibnamefont{Shimizu}}, \bibnamefont{and}
  \bibinfo{author}{\bibfnamefont{T.}~\bibnamefont{Kasuya}},
  \bibinfo{journal}{J. Phys. Soc. Japan} \textbf{\bibinfo{volume}{58}},
  \bibinfo{pages}{3666} (\bibinfo{year}{1989}).

\bibitem{CHZ94}
\bibinfo{author}{\bibfnamefont{T.~A.} \bibnamefont{Costi}},
  \bibinfo{author}{\bibfnamefont{A.}~\bibnamefont{Hewson}}, \bibnamefont{and}
  \bibinfo{author}{\bibfnamefont{V.}~\bibnamefont{Zlati\'c}},
  \bibinfo{journal}{J. Phys.: Condens. Matter} \textbf{\bibinfo{volume}{6}},
  \bibinfo{pages}{2519} (\bibinfo{year}{1994}).

\bibitem{xxx}
In Ref.~\onlinecite{MeZ71} Mezei and Zawadowski derived analytical
expressions for the Kondo-model LDOS for different types of scattering
in terms of an energy-dependent 
phase shift. However, they made 
explicit statements on only 
one energy point, for which they could refer to a definite value of the
phase shift.

\bibitem{CLE93}
\bibinfo{author}{\bibfnamefont{M.~F.} \bibnamefont{Crommie}},
  \bibinfo{author}{\bibfnamefont{C.~P.} \bibnamefont{Lutz}}, \bibnamefont{and}
  \bibinfo{author}{\bibfnamefont{D.~M.} \bibnamefont{Eigler}},
  \bibinfo{journal}{Phys. Rev. B} \textbf{\bibinfo{volume}{48}},
  \bibinfo{pages}{2851} (\bibinfo{year}{1993}).

\bibitem{LSBD98}
\bibinfo{author}{\bibfnamefont{J.}~\bibnamefont{Li}},
  \bibinfo{author}{\bibfnamefont{W.-D.} \bibnamefont{Schneider}},
  \bibinfo{author}{\bibfnamefont{R.}~\bibnamefont{Berndt}}, \bibnamefont{and}
  \bibinfo{author}{\bibfnamefont{B.}~\bibnamefont{Delley}},
  \bibinfo{journal}{Phys. Rev. Lett.} \textbf{\bibinfo{volume}{80}},
  \bibinfo{pages}{2893} (\bibinfo{year}{1998}).

\bibitem{MCJCW98}
\bibinfo{author}{\bibfnamefont{V.}~\bibnamefont{Madhavan}},
  \bibinfo{author}{\bibfnamefont{W.}~\bibnamefont{Chen}},
  \bibinfo{author}{\bibfnamefont{T.}~\bibnamefont{Jamneala}},
  \bibinfo{author}{\bibfnamefont{M.~F.} \bibnamefont{Crommie}},
  \bibnamefont{and} \bibinfo{author}{\bibfnamefont{N.~S.}
  \bibnamefont{Wingreen}}, \bibinfo{journal}{Science}
  \textbf{\bibinfo{volume}{280}}, \bibinfo{pages}{567} (\bibinfo{year}{1998}).

\bibitem{MLE00}
\bibinfo{author}{\bibfnamefont{H.~C.} \bibnamefont{Manoharan}},
  \bibinfo{author}{\bibfnamefont{C.~P.} \bibnamefont{Lutz}}, \bibnamefont{and}
  \bibinfo{author}{\bibfnamefont{D.~M.} \bibnamefont{Eigler}},
  \bibinfo{journal}{Nature} \textbf{\bibinfo{volume}{403}},
  \bibinfo{pages}{512} (\bibinfo{year}{2000}).

\bibitem{Nag65}
\bibinfo{author}{\bibfnamefont{Y.}~\bibnamefont{Nagaoka}},
  \bibinfo{journal}{Phys. Rev.} \textbf{\bibinfo{volume}{138}},
  \bibinfo{pages}{A 1112} (\bibinfo{year}{1965}).

\bibitem{ZiM68}
\bibinfo{author}{\bibfnamefont{J.}~\bibnamefont{Zittartz}},
  \bibnamefont{and}
  \bibinfo{author}{\bibfnamefont{E.}~\bibnamefont{M\"uller-Hartmann}}
  \bibinfo{journal}{Z. Phys.} \textbf{\bibinfo{volume}{212}},
  \bibinfo{pages}{380} (\bibinfo{year}{1968}).

\bibitem{NoM84}
\bibinfo{author}{\bibfnamefont{W.}~\bibnamefont{Nolting}} \bibnamefont{and}
  \bibinfo{author}{\bibfnamefont{M.}~\bibnamefont{Matlak}},
  \bibinfo{journal}{phys. stat. sol. (b)} \textbf{\bibinfo{volume}{123}},
  \bibinfo{pages}{155} (\bibinfo{year}{1984}).

\bibitem{HiF86}
\bibinfo{author}{\bibfnamefont{J.~E.} \bibnamefont{Hirsch}} \bibnamefont{and}
  \bibinfo{author}{\bibfnamefont{R.~M.} \bibnamefont{Fye}},
  \bibinfo{journal}{Phys. Rev. Lett.} \textbf{\bibinfo{volume}{56}},
  \bibinfo{pages}{2521} (\bibinfo{year}{1986}).

\bibitem{MeZ71}
\bibinfo{author}{\bibfnamefont{F.}~\bibnamefont{Mezei}} \bibnamefont{and}
  \bibinfo{author}{\bibfnamefont{A.}~\bibnamefont{Zawadowski}},
  \bibinfo{journal}{Phys. Rev. B} \textbf{\bibinfo{volume}{3}},
  \bibinfo{pages}{167} (\bibinfo{year}{1971}).

\bibitem{And61}
\bibinfo{author}{\bibfnamefont{P.~W.} \bibnamefont{Anderson}},
  \bibinfo{journal}{Phys. Rev.} \textbf{\bibinfo{volume}{124}},
  \bibinfo{pages}{41} (\bibinfo{year}{1961}).

\bibitem{ScW66}
\bibinfo{author}{\bibfnamefont{J.~R.} \bibnamefont{Schrieffer}}
  \bibnamefont{and} \bibinfo{author}{\bibfnamefont{P.~A.} \bibnamefont{Wolff}},
  \bibinfo{journal}{Phys. Rev.} \textbf{\bibinfo{volume}{149}},
  \bibinfo{pages}{491} (\bibinfo{year}{1966}).

\bibitem{SoS90}
\bibinfo{author}{\bibfnamefont{R.}~\bibnamefont{Sollie}} \bibnamefont{and}
  \bibinfo{author}{\bibfnamefont{P.}~\bibnamefont{Schlottmann}},
  \bibinfo{journal}{Phys. Rev. B} \textbf{\bibinfo{volume}{42}},
  \bibinfo{pages}{6099} (\bibinfo{year}{1990}).

\bibitem{ScH00}
\bibinfo{author}{\bibfnamefont{A.}~\bibnamefont{Schiller}} \bibnamefont{and}
  \bibinfo{author}{\bibfnamefont{S.}~\bibnamefont{Hershfield}},
  \bibinfo{journal}{Phys. Rev. B} \textbf{\bibinfo{volume}{61}},
  \bibinfo{pages}{9036} (\bibinfo{year}{2000}).

\bibitem{UKSZ00}
\bibinfo{author}{\bibfnamefont{O.}~\bibnamefont{\'Ujs\'aghy}},
  \bibinfo{author}{\bibfnamefont{J.}~\bibnamefont{Kroha}},
  \bibinfo{author}{\bibfnamefont{L.}~\bibnamefont{Szunyogh}}, \bibnamefont{and}
  \bibinfo{author}{\bibfnamefont{A.}~\bibnamefont{Zawadowski}},
  \bibinfo{journal}{Phys. Rev. Lett.} \textbf{\bibinfo{volume}{85}},
  \bibinfo{pages}{2557} (\bibinfo{year}{2000}).

\bibitem{YoY70}
\bibinfo{author}{\bibfnamefont{K.}~\bibnamefont{Yosida}} \bibnamefont{and}
  \bibinfo{author}{\bibfnamefont{K.}~\bibnamefont{Yamada}},
  \bibinfo{journal}{Suppl. Prog. Theor. Phys.} \textbf{\bibinfo{volume}{46}},
  \bibinfo{pages}{244} (\bibinfo{year}{1970}).

\bibitem{ZlH83}
\bibinfo{author}{\bibfnamefont{V.}~\bibnamefont{Zlati{\'c}}} \bibnamefont{and}
  \bibinfo{author}{\bibfnamefont{B.}~\bibnamefont{Horvati{\'c}}},
  \bibinfo{journal}{Phys. Rev. B} \textbf{\bibinfo{volume}{28}},
  \bibinfo{pages}{6904} (\bibinfo{year}{1983}).

\bibitem{MWPN99}
\bibinfo{author}{\bibfnamefont{D.}~\bibnamefont{Meyer}},
  \bibinfo{author}{\bibfnamefont{T.}~\bibnamefont{Wegner}},
  \bibinfo{author}{\bibfnamefont{M.}~\bibnamefont{Potthoff}}, \bibnamefont{and}
  \bibinfo{author}{\bibfnamefont{W.}~\bibnamefont{Nolting}},
  \bibinfo{journal}{Physica B} \textbf{\bibinfo{volume}{270}},
  \bibinfo{pages}{225} (\bibinfo{year}{1999}).

\bibitem{PWN97}
\bibinfo{author}{\bibfnamefont{M.}~\bibnamefont{Potthoff}},
  \bibinfo{author}{\bibfnamefont{T.}~\bibnamefont{Wegner}}, \bibnamefont{and}
  \bibinfo{author}{\bibfnamefont{W.}~\bibnamefont{Nolting}},
  \bibinfo{journal}{Phys. Rev. B}
  \textbf{\bibinfo{volume}{55}}(\bibinfo{number}{24}), \bibinfo{pages}{16132}
  (\bibinfo{year}{1997}).

\bibitem{MeN00Kon}
\bibinfo{author}{\bibfnamefont{D.}~\bibnamefont{Meyer}} \bibnamefont{and}
  \bibinfo{author}{\bibfnamefont{W.}~\bibnamefont{Nolting}},
  \bibinfo{journal}{Phys. Rev. B} \textbf{\bibinfo{volume}{61}},
  \bibinfo{pages}{13465} (\bibinfo{year}{2000}).

\bibitem{MeN00Dyn}
\bibinfo{author}{\bibfnamefont{D.}~\bibnamefont{Meyer}} \bibnamefont{and}
  \bibinfo{author}{\bibfnamefont{W.}~\bibnamefont{Nolting}},
  \bibinfo{journal}{Phys. Rev. B} \textbf{\bibinfo{volume}{62}},
  \bibinfo{pages}{5657} (\bibinfo{year}{2000}).

\bibitem{Fan61}
\bibinfo{author}{\bibfnamefont{U.}~\bibnamefont{Fano}}, \bibinfo{journal}{Phys.
  Rev.} \textbf{\bibinfo{volume}{124}}, \bibinfo{pages}{1866}
  (\bibinfo{year}{1961}).

\bibitem{RCMM75}
\bibinfo{author}{\bibfnamefont{L.~W.} \bibnamefont{Roeland}},
  \bibinfo{author}{\bibfnamefont{G.~J.} \bibnamefont{Cock}},
  \bibinfo{author}{\bibfnamefont{F.~A.} \bibnamefont{Muller}},
  \bibinfo{author}{\bibfnamefont{C.~A.} \bibnamefont{Moleman}},
  \bibinfo{author}{\bibfnamefont{K.~A.~M.} \bibnamefont{Mc\hspace{0.5em}Ewen}},
  \bibinfo{author}{\bibfnamefont{R.~C.} \bibnamefont{Jordan}},
  \bibnamefont{and} \bibinfo{author}{\bibfnamefont{D.~W.} \bibnamefont{Jones}},
  \bibinfo{journal}{J. Phys. F} \textbf{\bibinfo{volume}{5}},
  \bibinfo{pages}{L233} (\bibinfo{year}{1975}).

\bibitem{EvG68}
\bibinfo{author}{\bibfnamefont{H.~U.} \bibnamefont{Everts}} \bibnamefont{and}
  \bibinfo{author}{\bibfnamefont{B.~N.} \bibnamefont{Ganguly}},
  \bibinfo{journal}{Phys. Rev.} \textbf{\bibinfo{volume}{174}},
  \bibinfo{pages}{594} (\bibinfo{year}{1968}).

\bibitem{SZH89}
\bibinfo{author}{\bibfnamefont{D.}~\bibnamefont{\v{S}ok\v{c}evi\'c}},
  \bibinfo{author}{\bibfnamefont{V.}~\bibnamefont{Zlati\'c}}, \bibnamefont{and}
  \bibinfo{author}{\bibfnamefont{B.}~\bibnamefont{Horvati\'c}},
  \bibinfo{journal}{Phys. Rev. B} \textbf{\bibinfo{volume}{39}},
  \bibinfo{pages}{603} (\bibinfo{year}{1989}).

\end{thebibliography}
\end{document}